\documentclass{emulateapj}
\usepackage{natbib}
\usepackage{multirow}
\usepackage{color}

\begin{document}
\title{Investigating the relation between sunspots and umbral dots}

\author{Rahul Yadav\altaffilmark{1}, Rohan E. Louis\altaffilmark{2,3} and
Shibu K. Mathew\altaffilmark{1}}

\altaffiltext{1}{Udaipur Solar Observatory, Physical Research Laboratory
                 Dewali, Badi Road, Udaipur,
	       	     Rajasthan - 313004, India}

\altaffiltext{2}{Center of Excellence in Space Sciences India, Indian Institute 
                 of Science Education and Research Kolkata, Mohanpur--741246, 
                 West Bengal, India}
\altaffiltext{3}{Instituto de Astrof\'isica de Canarias, E-38200 La Laguna, Tenerife, Spain}

\email{rahuly@prl.res.in}

%%%%%%%%%%%%%%%%%%%%%%%%%%%%%%%%%%%%%%%%%%%%%%%%%%%
%%% Abstract 
\begin{abstract}
Umbral dots (UDs) are transient, bright features observed in the umbral region of 
a sunspot. We study the physical properties of UDs observed in sunspots of different 
sizes. The aim of our study is to relate the physical properties of umbral dots with 
the large-scale properties of sunspots. For this purpose, we analyze high-resolution 
G-band images of 42 sunspots observed by {\em Hinode}/SOT, located close to disk center. 
The images were corrected for instrumental stray-light and restored with the modeled 
PSF. An automated multi-level tracking algorithm was employed to identify the UDs located 
in selected G-band images. Furthermore, we employed HMI/SDO, limb-darkening corrected, 
full disk continuum images to estimate the sunspot phase and epoch for the selected 
sunspots. The number of UDs identified in different umbrae exhibits a linear relation 
with the umbral size. The observed filling factor ranges from 3\% to 7\% and increases 
with the mean umbral intensity. Moreover, the filling factor shows a decreasing trend 
with the umbral size. We also found that the observed mean and maximum 
intensities of UDs are correlated with the mean umbral intensity. However, we do not 
find any significant relationship between the mean (and maximum) intensity 
and effective diameter of umbral dots with the sunspot area, epoch, and decay rate. 
We suggest that this lack of relation could either be due to the distinct transition 
of spatial scales associated with overturning convection in the umbra or the shallow 
depth associated with umbral dots, or both the above.
\end{abstract}
%%%%%%%%%%%%%%%%%%%%%%%%%%%%%%%%%%%%%%%%%%%%%%%%%%%
%% Keywords

%-------------------------------------------------
%%%%%%%%%%%%%%%%%%%%%%%%%%%%%%%%%%%%%%%%%%%%%%%%%%%
%% Sections
%

\section{Introduction}
Umbral dots (UDs) are small, bright features observed in sunspot umbrae and pores. 
They cover only 3--10\% of the umbral area and contribute 10--20\% of its brightness 
\citep{1993ApJ...415..832S,2012ApJ...757...49W}. 
It has been suggested that umbral dots, light bridges etc. 
play a vital role in the energy balance of sunspots \citep{2003A&ARv..11..153S}.
While the strong magnetic field in the umbra suppresses energy transport by convection, 
some form of energy transport must be required to explain the observed umbral brightness. 

The nature of UDs has been described in a number of models. The cluster 
model of \cite{1979ApJ...234..333P} proposes that the umbral magnetic field is gappy, 
allowing field-free plasma to transport heat. An UD would represent 
the tip of such a field-free intrusion. The monolithic flux tube model of 
\cite{2002AN....323..371W} considers a sunspot as a collection of uniform 
vertically thin columns, and UDs as a natural result of the overstable oscillatory 
convection, which is the preferred mode just below the photosphere.
Simulations by \cite{2006ApJ...641L..73S} show that UDs are the result of 
narrow, upflowing, convective plumes with adjacent downflows.

Based on their location, UDs are classified as central umbral dots (CUDs) and peripheral 
umbral dots (PUDs) \citep{1986A&A...156..347G}. CUDs appear in the inner regions of the 
umbra, whereas PUDs are located near the umbra-penumbra boundary. The size of UDs ranges 
from {\color{magenta}}180--300~km and their intensity ranges from about 0.2 to 0.7 times 
the quiet Sun (QS) intensity at visible wavelengths 
\citep{2007PASJ...59S.585K,1997A&A...328..682S,2012ApJ...752..109L}. 
The distribution of UDs in the umbra is not uniform, \citet{1997A&A...328..682S} reported 
that larger, long-lived UDs are seen in regions of enhanced umbral background intensity. 
\citet{2009ApJ...702.1048W} also reported that UDs are likely to appear in regions where 
the magnetic field is weaker and inclined, whereas they tend to disappear in locations where 
the field is stronger and vertical. In addition, their study shows that the lifetimes and 
sizes of UDs are almost constant, regardless of the magnetic field strength. 
  
The physical properties of UDs have been extensively studied by several authors
\citep{1997A&A...328..682S, 2008A&A...492..233R,2011SoPh..270...75H,2012ApJ...757...49W,
2012ApJ...752..109L}, but they are primarily confined to time sequence observations of 
single spots or spots in a single active region (AR). \citet{2014PASJ...66S...1W} 
investigated the spatial distribution of UDs in several sunspots using {\em Hinode} 
observations. The results showed that UDs became more clustered in the latter phase 
of sunspots. If UDs are driven by small-scale magnetoconvection in umbrae, then
the sub-photospheric convective flows could influence the properties of UDs observed 
at the photosphere. The motivation of this article is to investigate if the 
macro-properties of sunspots, namely, area, umbral fill fraction, decay rate, and 
phase, have any bearing on the physical characteristics of UDs, specifically, 
intensity and size. To that extent, we combine observations from {\em Hinode} 
and and SDO/HMI to determine the properties of UDs and their host sunspots. The 
article is organized in the following manner. The observations and data analysis 
are described in Sect.~\ref{data}. The algorithm used for the identification of UDs 
is discussed in Sect.~\ref{mlt}. In Sect.~\ref{res}, we present our results. The 
discussion and conclusions are presented in Sects.~\ref{discuss} and \ref{conclu}, 
respectively. 

\begin{figure}[!t]
\centering
\includegraphics[width=0.45\textwidth]{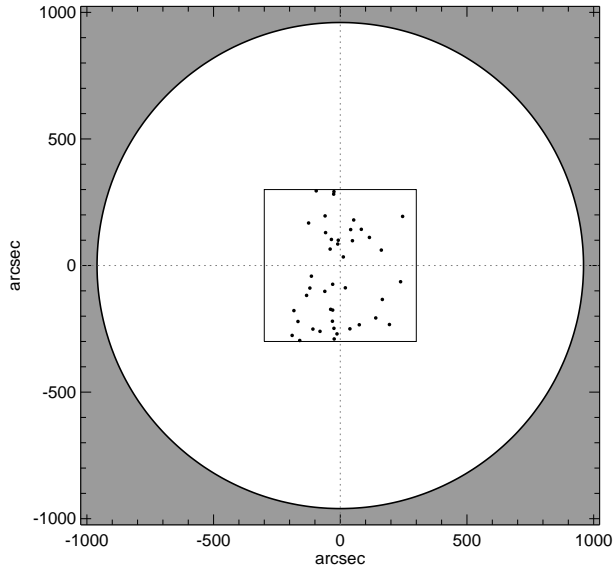}
\caption{The filled black circles depict the locations of the selected sunspots during 
{\em Hinode} observations on the artificial solar disk. The square represents the 
boundary of $\pm$5 arc-minute patch from the center of the solar disk.}
\label{spot-position}
\end{figure}

\section{Observations and data analysis}
\label{data}
In order to study the physical properties of UDs in different sunspots, we employed 
high resolution G-band images acquired by the Broadband Filter Imager 
(BFI) of the Solar Optical Telescope \citep[SOT;][]{2008SoPh..249..167T} on-board 
{\em Hinode} \citep{2007SoPh..243....3K}. The primary criterion for selecting the data 
was the proximity of sunspots to disc center to reduce the projection effects on the 
physical properties of UDs. Sunspots with heliocentric 
angles $\Theta \leqslant 25^\circ$ ($\mu \geqslant 0.9$) were chosen
for the study. We carefully examined the G-band images of fully evolved sunspots 
with well developed penumbrae from {\em Hinode} observations that were acquired 
between 2013 January to 2014 December. During these two years, we found 42 sunspots 
that met our selection criterion. This period coincides with the maximum phase of 
solar cycle 24. Out of 42 sunspots, 18 were 
located in the northern hemisphere (Fig.~\ref{spot-position}). 
The details of the chosen sunspots are listed in Table~\ref{noaalist}\footnote{The 
location of sunspots on the solar disk were identified using 
the following resources: https://helioviewer.org/ and https://www.solarmonitor.org/}.
All selected {\em Hinode} images were corrected for dark current, flat-field, and 
bad-pixels using routines available in the {\em Hinode} SolarSoft package. 
The spatial sampling for a majority of the {\em Hinode} G-band images was 
0\farcs22/pixel and 0\farcs11/pixel for 6 cases (Table \ref{noaalist}).

Full-disk, limb-darkening-removed, continuum images from the Helioseismic and 
Magnetic Imager \citep[HMI;][]{2012SoPh..275..207S}, on-board 
the Solar Dynamics Observatory \citep[SDO;][]{2012SoPh..275....3P}, were employed 
to estimate the epoch, growth and decay-rate of the selected spots. The HMI images 
are 4096~$\times$~4096 pixels in size, with a spatial sampling of around 0\farcs5/pixel
and a time cadence of 6~hr\footnote{Data are available at http://jsoc.stanford.edu/}.

\begin{table*}[!h]
\centering
\caption{Details of the ARs analyzed in the study. The effective umbral diameter 
(D$_\mathrm{umb}$), and the normalized mean umbral intensity (I$_\mathrm{umb}$) are given in the 
seventh, and eighth columns, respectively.}
\label{noaalist}
\resizebox{0.65\textwidth}{!}{%
\begin{tabular}{lcccrrrr}
\hline
Spot          & NOAA    & Date       & Time     & xpos     & ypos     & D$_\mathrm{umb}$ & I$_\mathrm{umb}$ \\
No.           &(AR $\#$)& (yy/mm/dd) & (UT)     & (\arcsec)& (\arcsec) & (\arcsec) &  \\
\hline
      1$_a$   &   11692 & 2013/01/14 & 04:17:42 &  -87 &  211 & 12.7 & 0.16\\
      2$_b$   &   11765 & 2013/06/07 & 07:31:31 &    9 &  115 & 10.8 & 0.18\\
      3$_b$   &   11785 & 2013/07/07 & 21:21:30 &   -4 & -260 & 16.3 & 0.11\\
      4$_a$   &   11809 & 2013/08/06 & 04:39:06 &   -3 &   59 &  8.8 & 0.17\\
      5$^L_a$ &   11861 & 2013/10/12 & 18:10:02 &   45 & -234 & 16.6 & 0.12\\
      6$^F_a$ &   11861 & 2013/10/12 & 18:10:02 &  -55 & -248 & 14.5 & 0.14\\
      7$_a$   &   11884 & 2013/11/07 & 17:32:10 & -254 & -274 & 19.5 & 0.09\\
      8$_b$   &   11890 & 2013/11/19 & 10:43:02 &  181 &   30 & 34.3 & 0.08\\
      9$^L_a$ &   11921 & 2013/12/15 & 17:41:38 &  -10 &   85 & 19.7 & 0.10\\
     10$^F_a$ &   11921 & 2013/12/15 & 17:41:38 &  -10 &   85 & 14.1 & 0.11\\
     11$_b$   &   11934 & 2013/12/26 & 10:28:02 &   58 & -273 & 13.3 & 0.14\\
     12$^L_a$ &   11944 & 2014/01/07 & 08:44:58 &  -74 &  -65 & 40.0 & 0.09\\
     13$^F_a$ &   11944 & 2014/01/07 & 08:44:58 & -180 &  -65 & 20.0 & 0.14\\
     14$_a$   &   11959 & 2014/01/23 & 10:48:31 & -179 & -214 & 18.8 & 0.14\\
     15$_a$   &   11960 & 2014/01/24 & 14:49:02 &   10 & -332 & 16.2 & 0.12\\
     16$^L_a$ &   11967 & 2014/02/03 & 06:08:14 &   15 & -147 & 37.9 & 0.10\\
     17$^F_a$ &   11967 & 2014/02/03 & 06:08:14 &   15 & -147 & 30.6 & 0.11\\
     18$_b$   &   11974 & 2014/02/10 & 23:16:36 & -210 & -110 & 13.1 & 0.14\\
     19$_a$   &   11990 & 2014/03/03 & 01:45:03 &   50 & -138 & 14.7 & 0.13\\
     20$_a$   &   11991 & 2014/03/03 & 06:55:01 & -108 & -251 &  9.1 & 0.16\\
     21$_a$   &   12002 & 2014/03/12 & 13:56:04 & -254 & -225 & 14.7 & 0.15\\
     22$_a$   &   12005 & 2014/03/18 & 00:35:00 &  -95 &  285 & 18.7 & 0.12\\
     23$_a$   &   12014 & 2014/03/26 & 11:29:57 &  166 & -134 & 15.0 & 0.12\\
     24$_a$   &   12027 & 2014/04/06 & 04:16:59 &  -43 &  282 & 16.3 & 0.12\\
     25$_a$   &   12032 & 2014/04/13 & 16:14:59 &  -61 &  282 & 15.6 & 0.11\\
     26$^L_a$ &   12056 & 2014/05/11 & 18:50:57 &  -35 &  103 & 12.4 & 0.14\\
     27$^F_a$ &   12056 & 2014/05/11 & 18:50:57 & -125 &  168 & 13.0 & 0.13\\
     28$_a$   &   12080 & 2014/06/08 & 03:08:58 &  -54 & -243 & 13.1 & 0.11\\
     29$_a$   &   12096 & 2014/06/28 & 18:11:58 &  127 &   33 &  8.7 & 0.17\\
     30       &   12104 & 2014/07/04 & 20:25:26 &  -13 & -270 & 12.1 & 0.15\\
     31       &   12104 & 2014/07/04 & 20:25:26 &  -31 & -220 & 10.0 & 0.15\\
     32$_a$   &   12121 & 2014/07/27 & 21:25:29 & -114 &  -42 & 13.6 & 0.14\\
     33$_a$   &   12135 & 2014/08/11 & 02:52:00 &  -93 &   69 & 14.3 & 0.14\\
     34$_a$   &   12146 & 2014/08/22 & 18:28:41 &  -26 &   -9 & 14.6 & 0.12\\
     35$_a$   &   12151 & 2014/08/30 & 08:07:49 &  165 & -287 & 14.2 & 0.13\\
     36$_a$   &   12158 & 2014/09/11 & 10:46:00 &   66 &   83 & 23.2 & 0.12\\
     37$^L_a$ &   12172 & 2014/09/26 & 07:32:08 &  -24 & -290 & 18.7 & 0.10\\
     38$^F_a$ &   12172 & 2014/09/26 & 07:32:08 & -190 & -276 & 18.1 & 0.14\\
     39$_a$   &   12178 & 2014/10/03 & 07:55:27 &  -38 & -173 & 14.4 & 0.12\\
     40$_a$   &   12205 & 2014/11/10 & 08:26:58 &  -10 &  162 & 12.6 & 0.13\\
     41$_a$   &   12216 & 2014/11/26 & 06:00:58 &   10 & -287 & 17.3 & 0.11\\
     42$_b$   &   12227 & 2014/12/09 & 14:04:35 &  171 & -108 & 13.3 & 0.15\\
\hline
\end{tabular}
}
\begin{flushleft}
\textbf{Notes.} (L) Leading spot. (F) Following spot. \\(a) spatial sampling: 
0\farcs22/pixel. (b) spatial sampling: 0\farcs11/pixel.
\end{flushleft}
\end{table*}

\subsection{Instrumental stray-light correction}
\label{stray}
Space-based observations are free from {\em seeing} effects, but the optical 
quality of an instrument degrades over a period of time resulting in decreased 
sensitivity and image contrast \citep{2007A&A...465..291M}. Instrumental scattered 
light can greatly influence the physical properties of sunspot fine structure 
\citep{2012ApJ...752..109L}. All G-band images were corrected for instrumental 
stray-light using the point-spread function (PSF) derived by 
\cite{2009A&A...501L..19M}. They determined the PSF of the broad-band images of 
the SOT by analyzing the transit of Mercury observed on 2006 November 6. The PSF 
was modelled as a combination of four Gaussians with different widths and weights 
(Table~1 of \cite{2007A&A...465..291M}). Restoration of the images was carried out 
by performing a deconvolution using the maximum likelihood approach 
\citep{1972JOSA...62...55R,1974AJ.....79..745L} available in the IDL Astrolib 
package. This method iteratively updates the current estimate of the image by the
product of the previous deconvolution and the correlation between the re-convolution 
of the subsequent image and the PSF. The G-band images were normalized 
to the average quiet Sun intensity (I$_\mathrm{{QS}}$) over a $100\times100$~pixel area far 
away from the spot. Bright points in this patch exceeding 1.6I$_{\mathrm{QS}}$ were ignored while computing the mean intensity.

\begin{figure}[!h]
\centering
\includegraphics[width=0.47\textwidth]{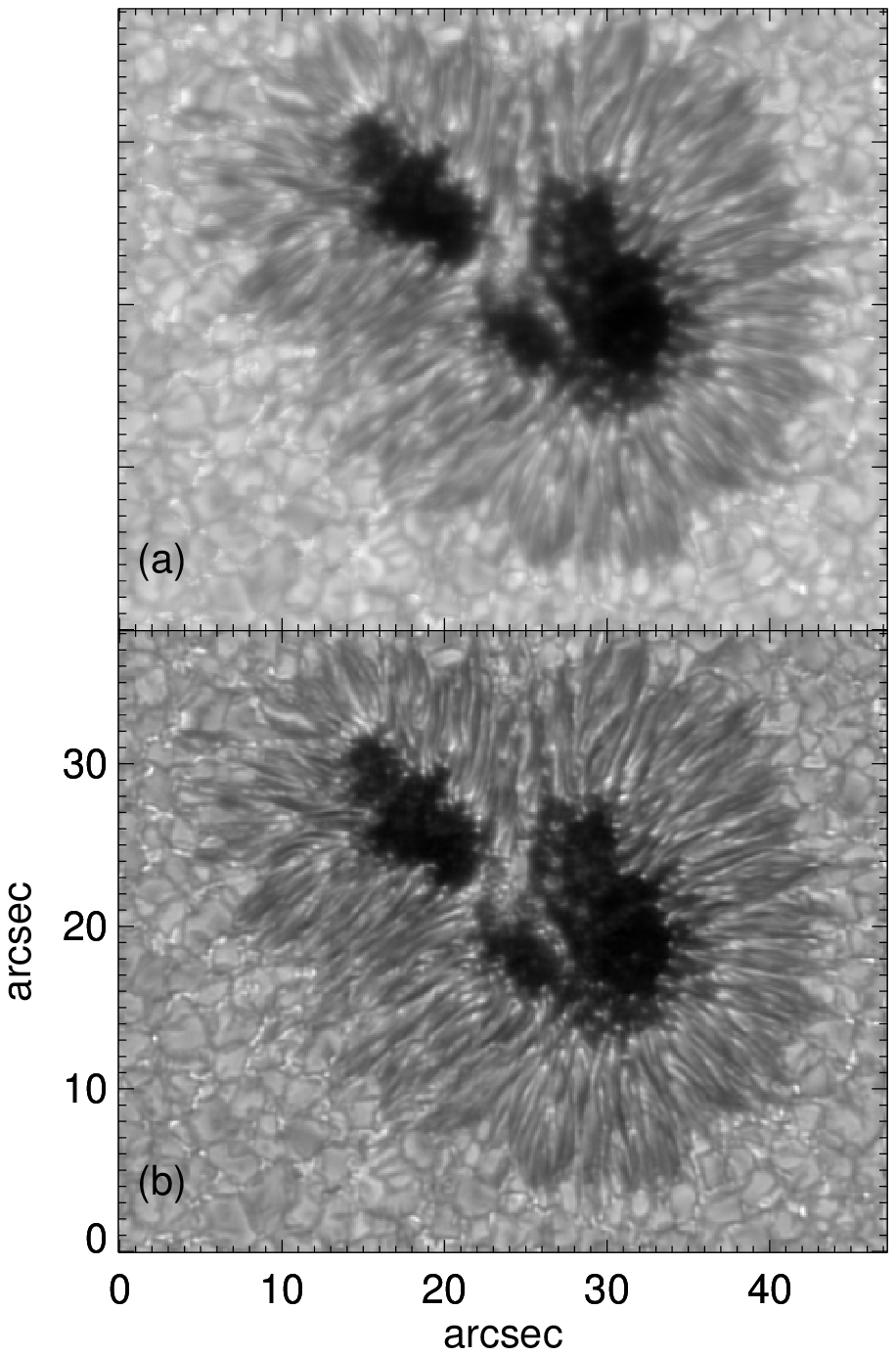}
\caption{Sunspot in NOAA AR~12227 shown before (a) and after (b) stray-light 
correction. Both images in the top and bottom panels have the same intensity scale.}
\label{stray-compare}
\end{figure}

Figure~\ref{stray-compare} shows the improvement in image contrast after  
stray-light correction for a sunspot in NOAA AR~12227, observed on 2014 December 
9 by {\em Hinode}/SOT. 
%%%After stray-light correction, all the images were normalized 
%%%with the mean quiet-Sun intensity. 
The mean minimum umbral intensity in selected 
sunspots decreases from 0.096I$_\mathrm{QS}$ to 0.044I$_\mathrm{QS}$ and the mean contrast 
in the umbra increases from 0.4 to 0.51, after removal of stray-light. 
Hereafter, the uncorrected and stray-light corrected G-band images will be referred 
to as NC and SC images, respectively.

\subsection{Sunspot epoch, decay and growth rate} 
\label{epoch}
Full-disk HMI continuum images were used to determine the macro-properties of the
sunspots. Prior to estimating the sunspot area, the images were corrected 
for geometric foreshortening in the following manner. The position of each pixel 
on the solar disk was defined in terms of the position angle 
$\theta_{i}=\tan^{-1}(x_{i}/y_{i})$ and the radial distance 
$\rho_{i} =\sqrt{x^2_{i}+y^2_{i}}$, where $x_{i}$ and $y_{i}$ represent the 
position of $i^{th}$ pixel on the surface of the Sun measured from disc center. 
The heliographic coordinates (B and L), were calculated from the following 
equations:

\begin{equation}
B=\sin^{-1} \{\sin{B_{0}}\cos\rho + \cos{B_{0}}\sin\rho\cos(P-\theta_{i})\}
\end{equation}
\begin{equation}
L=\sin^{-1} \Big\{\frac{\sin\rho_{i}\cos(P-\theta_{i})}{ \cos{B}}\Big\} + L_{0}
\end{equation}
\begin{equation}
\rho = \sin^{-1}\Big\{\frac{\rho_{i}}{S}\Big\}-\rho_{i},
\end{equation}

where {\em S} is the radius of the Sun in arcsec and {\em P} is the equatorial 
horizontal parallax angle. Once the heliographic coordinates for each pixel 
were known they were transferred to a two-dimensional (2D) image defined by 
latitude and longitude. This process was carried out on all HMI continuum 
images for the selected sunspots. The HMI filtergrams were normalized with the 
QS intensity at disc centre.
 
\begin{figure}[!h]
\centering
\includegraphics[width=0.5\textwidth]{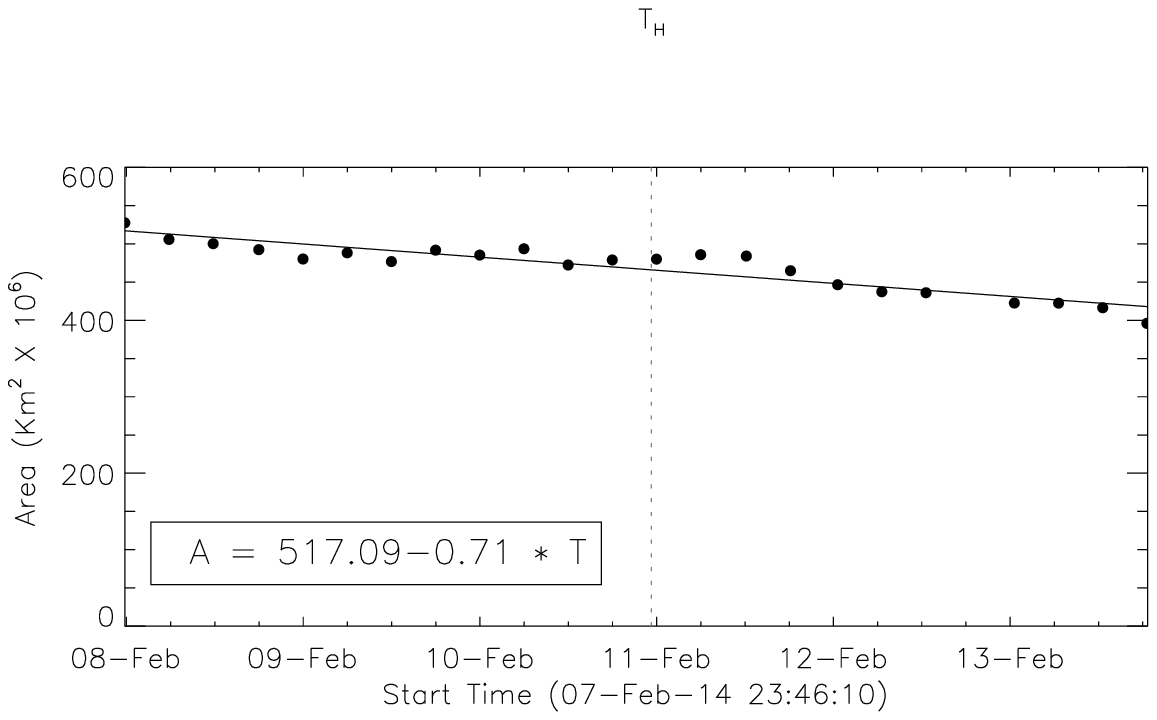}
\caption{Temporal evolution of an umbral area.. The filled circles represents the area 
of the umbra (NOAA AR~11974) and the solid line is a linear fit. A grey 
dotted line indicates the {\em Hinode} observing time (T$_\mathrm{H}$) of the same sunspot.}
\label{decay-plot}
\end{figure}

Before calculating the sunspot area, the intensity corresponding to the umbra-penumbra 
boundary and the penumbra-QS boundary was determined using the cumulative intensity 
histogram method \citep{1997SoPh..175..197P}. The intensity threshold for the 
umbra-penumbra and penumbra--QS boundaries were estimated to be 0.5I$_\mathrm{QS}$ and 
0.9I$_\mathrm{QS}$, respectively. The variation of the spot area with time was used to determine 
the decay or growth rate using a linear fit \citep{2003GRL_chapman:GRL16419}. The area 
change of a sunspot as a function of time can be expressed as, 
$\mathrm{A(T) = m\times (T - T_0) + A_{0}}$, where $m$ yields the decay or growth rate and 
$A_{0}$ is the maximum area of a decaying spot (minimum in the case of a
growing spot). From the above expression, the time when the area of a decaying spot 
reduces to zero works out to be $\mathrm{T_{n} = A_0/m}$, where $\mathrm{T_{0}}$
is the time when the area of a decaying sunspot is maximum (minimum in case of 
growing spot) on the solar disk. In order to relate the macro-properties
of a sunspot with the physical characteristics of UDs, we determined the epoch of a decaying spot as the ratio of the {\em Hinode} observing time ($\mathrm{T_H}$) to the time when 
the spot area reduces to zero ($\mathrm{T_{n}}$), whereas for the growing spot it is defined as the ratio of the {\em Hinode} observing time ($\mathrm{T_H}$) to the time when 
the spot area is maximum. As an example, the decay rate of a sunspot in NOAA AR~11974 is shown in 
Fig.~\ref{decay-plot}. The area rate of change and epoch of the spots are 
described in the sixth and seventh columns of Table~\ref{udpara}, respectively. 

\section{Identification of UDs}
\label{mlt}
The first step in identifying UDs from the {\em Hinode} G-band images was to extract the umbral region. The SC images,
corresponding to a spatial sampling of 0$\farcs$11~pixel$^{-1}$, were smoothed by 
a 15~pixel~$\times$~15~pixel boxcar (a 7.5~pixel~$\times$~7.5~pixel boxcar
was used for spots with a coarser spatial sampling) and the umbra was extracted 
using the cumulative histogram method employed earlier for the HMI images. 
This yielded a value of 0.35I$_\mathrm{QS}$ corresponding to the umbra-penumbra 
boundary in the {\em Hinode} G-band images. The identification of UDs was carried 
out using a 2D multi-level tracking 
\citep[MLT;][]{2001SoPh..201...13B} algorithm that has been implemented by 
\cite{2008A&A...492..233R} and \cite{2012ApJ...752..109L}. This algorithm 
works in the following manner. The intensity range in the umbra 
is binned into several levels chosen by the user and the algorithm identifies UDs at 
each intensity level starting from the highest to the lowest intensity level. 
All UDs corresponding to the maximum intensity in the umbra are tagged uniquely. 
Then, the intensity threshold is reduced to the next lower level and the UDs are 
identified once again, with the previously identified structures retaining their 
tagged number. 
%%---%%UDs that are identified in this level are given a new number. 
This process continues until the last intensity level is reached. The number
of UDs detected increases with the number of intensity levels defined in MLT algorithm.
%%After carefully examining the number of UDs as a function of intensity levels, 
%%we set 25~levels for the SC images. 

In order to optimize the MLT algorithm for our dataset, 
the number of intensity levels was varied between 10 to 45 in steps of 5 and the 
corresponding values of the physical parameters were noted. We observed that although 
the number of UDs detected, increased by a factor of about 1.5 when the number of 
intensity levels changed from 25 to 45, the average value of the physical parameters 
remained unchanged and the overall variation for different sized-sunspots was nearly 
similar at and above 25 intensity levels. Furthermore, a visual inspection was carried 
out to verify if the algorithm identified all discernible UDs for a given number of 
levels. With 10 intensity levels, obvious UDs go undetected, while with 45 levels the 
algorithm primarily detects smaller and diffuse structures, whose inclusion does 
not alter the final statistical results. A final test was performed, with a smaller 
sample of sunspots, spanning 10\arcsec~to 40\arcsec~in diameter, and it was found 
that the final results were unaffected. This test was also carried out for different 
number of intensity levels and the outcome was similar to the one described above. 
These tests allowed us to finally select 25 intensity levels for the MLT routine.

\begin{figure}[!h]
\centering
\includegraphics[width=0.48\textwidth]{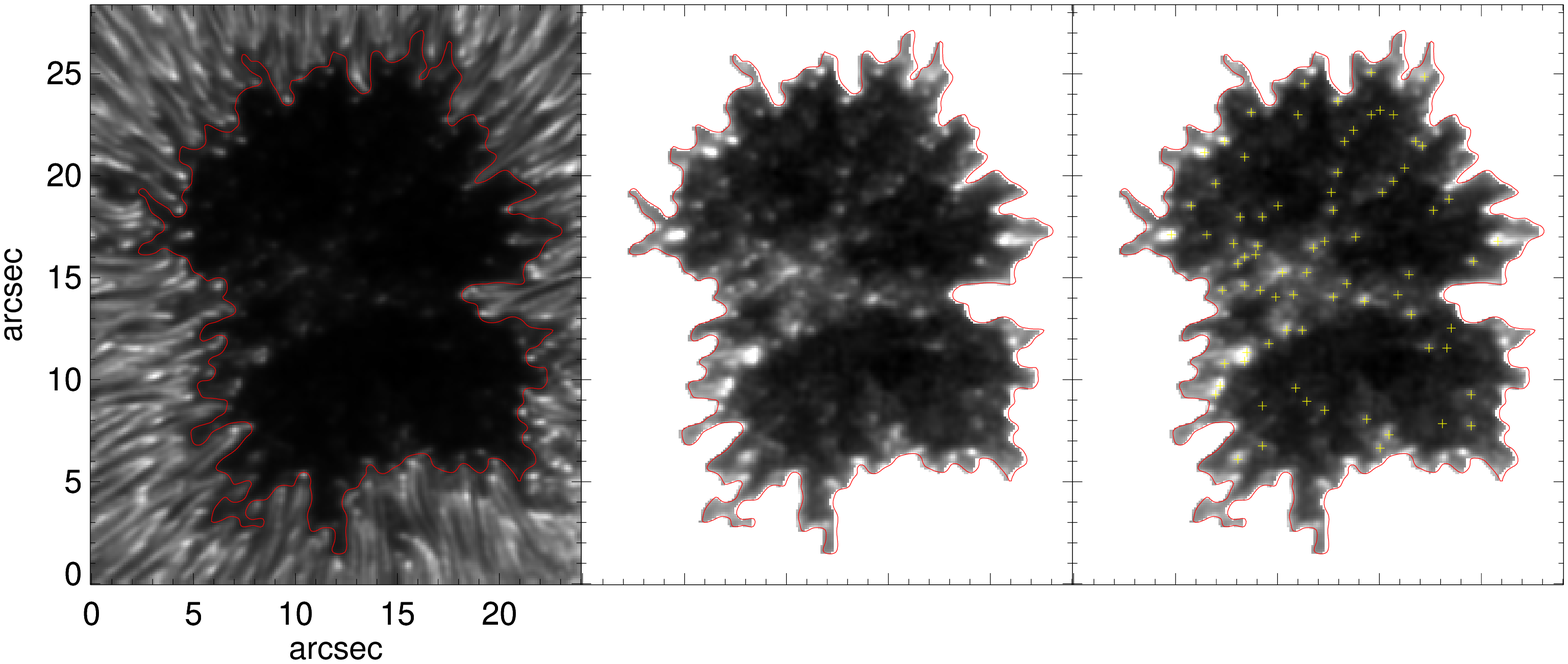}
\includegraphics[width=0.48\textwidth]{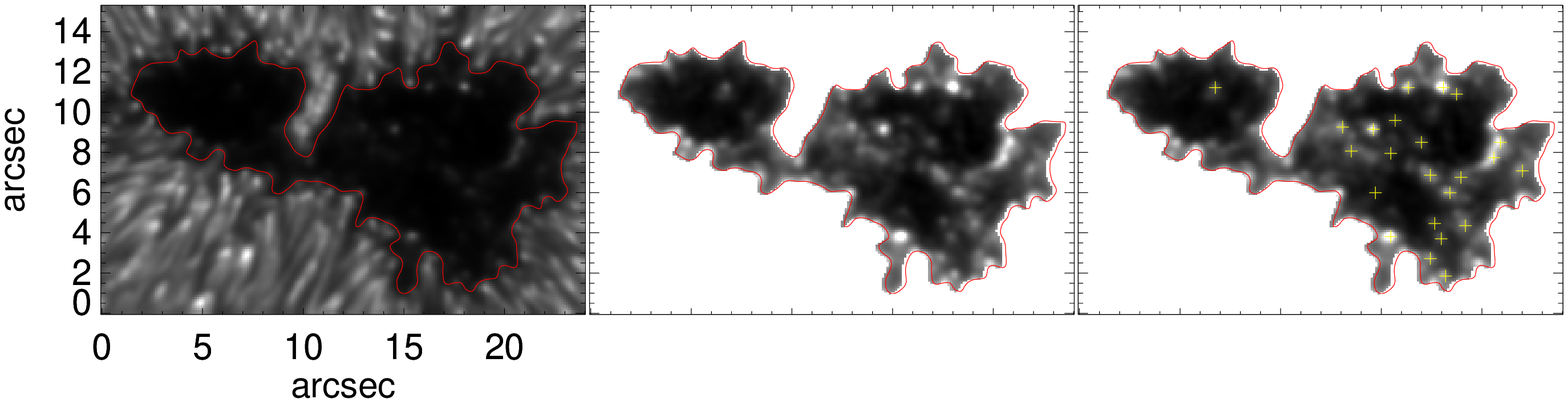}
\caption{Top panels: Stray-light corrected image of a 
sunspot observed in NOAA AR~11921 where the red contour outlines the 
umbra--penumbra boundary (left), the isolated umbra from the sunspot (middle), and
%%%%and the penumbra region of the sunspot has been masked to highlight the 
%%%%structuring in the umbra and the umbral intensity is enhanced to show 
%%%%the locations of dim UDs. 
the isolated umbra with the locations of the UDs indicated by yellow plus symbols
after employing the MLT algorithm with 25 intensity levels (right). Bottom panels: 
Same as top panels but for NOAA AR~11974.}
\label{mlt-demo}
\end{figure}

Once the UDs are identified by the MLT routine, the boundary of each UD is defined 
by a contour which corresponds to 50\% of its 
maximum and background intensity, i.e., (I$_\mathrm{max}$ + I$_\mathrm{bg}$)/2. The background 
umbral image is determined by smoothing the original image with a 
7~pixel~$\times$~7~pixel boxcar window. Figure~\ref{mlt-demo} shows the extraction 
of an umbra from a sunspot and the location of UDs identified using the MLT algorithm. 
 
\begin{table*}[!h]
\centering
\caption{Physical properties of UDs determined from the {\em Hinode} 
stray-light corrected images. Columns 2--6 correspond to the mean intensity
(I$_\mathrm{mean}$/I$_\mathrm{QS}$), maximum intensity (I$_\mathrm{max}$/I$_\mathrm{QS}$), the mean effective diameter 
(D$_\mathrm{eff}$), total number of umbral dots (UD $\#$), and the filling factor (ff), 
respectively. The seventh and eighth columns represent the area change in the selected 
sunspots and epoch, respectively which were estimated from \textit{HMI/SDO} 
continuum images. The values in the parentheses represent the $rms$ value.}
\label{udpara}
\resizebox{0.7\textwidth}{!}{%
\begin{tabular}{llllrlll}
\hline
Spot & I$_\mathrm{mean}$/I$_\mathrm{QS}$ & I$_\mathrm{max}$/I$_\mathrm{QS}$   & D$_\mathrm{eff}$& UD  & ff  & Area change & epoch \\
No.  &                     &                  & (\arcsec)    &$\#$ &(\%) & 10$^6$ $\times$  Km$^2$/day  \\
	 \hline
1  &  0.23 (0.09) &  0.26 (0.11) &  0.64 (0.14) &  24 &  6.3 &  10.63 (2.35)$^{\dagger}$  &      0.054\\
2  &  0.27 (0.07) &  0.30 (0.09) &  0.35 (0.07) &  52 &  5.5 & 113.06 (26.57)$^{\dagger}$ &      0.188\\
3  &  0.30 (0.18) &  0.36 (0.25) &  0.51 (0.14) &  33 &  3.4 & 110.04 (23.06)$^{\dagger}$ &      0.262\\
4  &  0.21 (0.07) &  0.24 (0.09) &  0.57 (0.11) &  15 &  6.5 &  36.44 (3.55)$^{\dagger}$  &      0.383\\
5  &  0.17 (0.09) &  0.20 (0.11) &  0.69 (0.16) &  33 &  6.0 &  68.07 (19.90)$^{\perp}$   &      0.507\\
6  &  0.22 (0.15) &  0.26 (0.19) &  0.64 (0.15) &  27 &  5.6 &  47.78 (4.06)$^{\dagger}$  &      0.087\\
7  &  0.18 (0.14) &  0.21 (0.18) &  0.68 (0.17) &  37 &  4.8 &  53.80 (10.97)$^{\dagger}$ &      0.097\\
8  &  0.21 (0.13) &  0.25 (0.15) &  0.55 (0.17) & 128 &  3.5 &   9.34 (4.66)$^{\perp}$    &      0.073\\
9  &  0.18 (0.06) &  0.21 (0.07) &  0.47 (0.15) &  77 &  4.5 &  57.99 (9.07)$^{\dagger}$  &      0.244\\
10 &  0.16 (0.06) &  0.19 (0.08) &  0.49 (0.17) &  49 &  6.2 &  95.77 (14.42)$^{\perp}$   &      0.521\\
11 &  0.24 (0.12) &  0.28 (0.15) &  0.46 (0.08) &  49 &  6.0 &  29.82 (3.22)$^{\dagger}$  &      0.225\\
12 &  0.18 (0.09) &  0.22 (0.12) &  0.83 (0.32) & 114 &  5.3 &  42.58 (3.19)$^{\dagger}$  &      0.101\\
13 &  0.20 (0.06) &  0.24 (0.08) &  0.69 (0.21) &  47 &  6.0 &  54.65 (10.42)$^{\dagger}$ &      0.046\\
14 &  0.22 (0.11) &  0.25 (0.14) &  0.50 (0.12) &  82 &  6.1 &  11.02 (1.80)$^{\dagger}$  &      0.039\\
15 &  0.22 (0.13) &  0.26 (0.16) &  0.66 (0.14) &  21 &  3.7 &  31.47 (2.64)$^{\dagger}$  &      0.132\\
16 &  0.22 (0.13) &  0.26 (0.16) &  0.74 (0.23) &  98 &  4.1 & 153.73 (49.73)$^{\dagger}$ &      0.034\\
17 &  0.22 (0.14) &  0.26 (0.18) &  0.70 (0.19) &  59 &  3.4 & 223.40 (28.32)$^{\perp}$   &      0.681\\
18 &  0.25 (0.14) &  0.29 (0.18) &  0.54 (0.14) &  23 &  4.1 &  17.14 (1.61)$^{\dagger}$  &      0.099\\
19 &  0.21 (0.10) &  0.24 (0.14) &  0.66 (0.15) &  24 &  5.1 &  61.01 (6.17)$^{\dagger}$  &      0.116\\
20 &  0.22 (0.10) &  0.24 (0.11) &  0.62 (0.18) &  10 &  5.1 &  69.52 (22.10)$^{\dagger}$ &      0.151\\
21 &  0.22 (0.15) &  0.25 (0.19) &  0.58 (0.13) &  27 &  4.4 &  92.84 (4.66)$^{\dagger}$  &      0.049\\
22 &  0.19 (0.14) &  0.22 (0.17) &  0.64 (0.22) &  41 &  5.3 &  31.56 (2.16)$^{\dagger}$  &      0.125\\
23 &  0.17 (0.11) &  0.19 (0.13) &  0.66 (0.16) &  28 &  5.7 &  37.49 (2.90)$^{\perp}$    &      0.572\\
24 &  0.15 (0.07) &  0.17 (0.10) &  0.70 (0.11) &  27 &  5.2 &  33.12 (2.45)$^{\dagger}$  &      0.122\\
25 &  0.14 (0.07) &  0.16 (0.09) &  0.70 (0.16) &  23 &  4.8 &  12.71 (1.87)$^{\dagger}$  &      0.048\\
26 &  0.19 (0.08) &  0.22 (0.09) &  0.57 (0.13) &  32 &  7.0 &  37.13 (2.35)$^{\dagger}$  &      0.170\\
27 &  0.14 (0.07) &  0.16 (0.07) &  0.68 (0.22) &  10 &  3.0 &  25.14 (1.70)$^{\dagger}$  &      0.123\\
28 &  0.17 (0.11) &  0.20 (0.14) &  0.69 (0.18) &  11 &  3.2 &  23.31 (2.3)$^{\perp}$     &      0.233\\
29 &  0.28 (0.09) &  0.32 (0.12) &  0.66 (0.24) &   8 &  5.1 &  55.84 (5.04)$^{\dagger}$  &      0.466\\
30 &  0.20 (0.08) &  0.22 (0.09) &  0.67 (0.16) &  22 &  7.2 &  73.44 (6.00)$^{\dagger}$  &      0.225\\
31 &  0.17 (0.08) &  0.19 (0.09) &  0.58 (0.15) &  13 &  4.5 &  86.37 (5.26)$^{\dagger}$  &      0.310\\
32 &  0.20 (0.09) &  0.22 (0.10) &  0.65 (0.13) &  25 &  5.9 &  29.72 (2.76)$^{\dagger}$  &      0.135\\
33 &  0.16 (0.06) &  0.18 (0.08) &  0.72 (0.14) &  21 &  5.5 &  11.96 (1.34)$^{\dagger}$  &      0.056\\
34 &  0.16 (0.07) &  0.19 (0.09) &  0.70 (0.17) &  23 &  5.6 &  44.36 (7.44)$^{\perp}$    &      0.635\\
35 &  0.17 (0.10) &  0.19 (0.12) &  0.75 (0.11) &  20 &  5.8 &  26.69 (2.24)$^{\dagger}$  &      0.132\\
36 &  0.19 (0.07) &  0.21 (0.08) &  0.66 (0.15) &  65 &  5.6 & 130.75 (23.23)$^{\dagger}$ &      0.302\\
37 &  0.11 (0.06) &  0.12 (0.07) &  0.61 (0.15) &  44 &  4.9 & 252.80 (38.66)$^{\dagger}$ &      0.023\\
38 &  0.20 (0.10) &  0.22 (0.12) &  0.65 (0.16) &  45 &  6.2 &  38.07 (5.59)$^{\dagger}$  &      0.061\\
39 &  0.29 (0.21) &  0.36 (0.28) &  0.84 (0.14) &  13 &  4.5 &  13.30 (7.51)$^{\dagger}$  &      0.042\\
40 &  0.21 (0.10) &  0.23 (0.12) &  0.65 (0.16) &  21 &  5.9 &  44.47 (12.48)$^{\dagger}$ &      0.243\\
41 &  0.15 (0.09) &  0.16 (0.10) &  0.67 (0.15) &  22 &  3.5 & 137.53 (23.02)$^{\dagger}$ &      0.299\\
42 &  0.25 (0.16) &  0.28 (0.19) &  0.38 (0.10) &  68 &  5.8 &  15.40 (1.49)$^{\dagger}$  &      0.098\\
\hline
\end{tabular}
}
\begin{flushleft}
\textbf{Notes.} ($\perp$) Growing spot. ($^{\dagger}$) Decaying spot.
\end{flushleft}
\end{table*}

For each UD, we determine its peak intensity and mean intensity over all 
pixels enclosed by the UD boundary.
We assume that the shape of each UD is circular, although they can be 
elliptical \citep{2012ApJ...745..163K}. The effective diameter for each 
identified UD was calculated as D$_\mathrm{eff}=\sqrt{4A/\pi}$, where $A$ represents 
the total number of pixels in an UD. Those structures which
had an area of greater than 2~pixels were considered for the analysis. We also
verified if selecting UDs on the basis of their proximity to the umbra-penumbra boundary
had any discernible effect on the final results and we found that neither an ingress
of the umbra-boundary, moving from 2 to 8~pixels inwards, nor excluding conspicuous bright
peripheral features altered the average value of the physical parameters chosen for our study.

For each sunspot, the mean of the following quantities pertaining to the UDs was 
determined, namely, maximum intensity, mean intensity, and effective diameter. 
Hereafter, the term intensity will be used to refer to the average of both 
the mean and maximum intensity of UDs, unless explicitly mentioned.
%%%%It is well-known that peripheral UDs are brighter than central 
%%%%UDs and tend to have physical propertied similar to bright penumbral features. In order
%%%%to check if the inclusion of such features bias the final result, we moved the 
%%%%umbra-penumbra boundary inwards by 8~pixels to exclude any UDs that might be similar to 
%%%%structures in the penumbra. An ingress of the umbra-penumbra boundary as descried above, 
%%%%does not alter the final results.}
   
\section{Results}
\label{res}
In this section, we present the physical properties of UDs observed in the umbra 
of 42 different sunspots, which are summarized in Table~\ref{udpara}.

\subsection{Umbral size, intensity and number of UDs}
The umbrae (D$_\mathrm{umb}$) from the 42 selected sunspots in the {\em Hinode} 
data set 
%%%were isolated using the intensity threshold approach described in 
%%%\cite{1997SoPh..175..197P}. The umbra--penumbra boundary was defined using an intensity
%%threshold of 0.35$I_{QS}$. We found that the umbral diameter 
varied from 8\arcsec~to 40\arcsec~ in diameter (see Table~\ref{noaalist}). However, the 
majority of sunspots had a diameter between 10\arcsec and 20\arcsec and only 4 
sunspots with a diameter between 30\arcsec and 40\arcsec.

\begin{figure}[!h]
\centering
\includegraphics[width=\columnwidth]{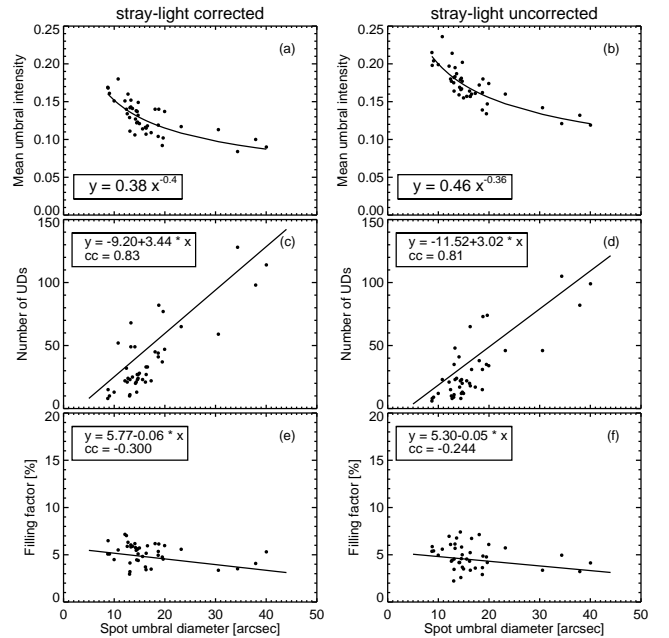}
\caption{Left panel: From top to bottom - variation of the mean umbral intensity, 
number of UDs, and filling factor as a function of the umbral diameter for the stray-light 
corrected images. Right panel: same as left but for the uncorrected data set. The solid 
line in the top panels represents a power law fit, while in the middle and bottom panels 
it corresponds to a linear fit. The linear correlation coefficient between the parameters
is indicated by `cc' in the inset.}
\label{size-para}
\end{figure}

Panel `a' of Figure~\ref{size-para} shows the relation between the mean intensity in 
the umbra and umbral diameter for the SC data set. A non-linear relation between the 
two quantities is clearly evident. The mean intensity in the umbra for all spots varies 
between 0.08--0.19I$_\mathrm{QS}$. However, in smaller umbrae (D$_\mathrm{umb}<20''$) the intensity 
decreases much faster than in larger umbrae. For D$_\mathrm{umb}>20''$ the intensity decreases 
slowly and linearly. In order to understand 
this behaviour, the observed distribution is fitted with linear and power law functions. 
However, we noticed that a power law fit describes this distribution better than a 
linear fit. The power law describing the non-linear trend is indicated by a solid line 
and the corresponding parameters are included in Table~\ref{table_fitparam}. It is also 
observed that smaller umbrae are brighter than the bigger ones by a factor of 1.7.
In order to verify that the observed trend between the mean umbral intensity and the 
umbral diameter is not an artifact from the PSF deconvolution, we plot the same for 
the uncorrected data set as well, which is shown in Panel `b' of Fig.~\ref{size-para}. 
The non-linear trend is clearly visible, although the intensities are much higher, 
ranging from 0.12--0.24I$_\mathrm{QS}$ which would be expected from the uncorrected data set. 
The plots also show that the slopes for the SC and NC data sets only differ by 10\%.

Panel `c' of Fig.~\ref{size-para} demonstrates the relation between the number of UDs 
(NUDs) and the umbral diameter for the SC images, and it is seen that the NUDs increase 
with the umbral diameter. To a first order, this scatter can be expressed with a linear 
fit between the two quantities. The same is observed with the NC images 
(Panel `d'), although the NUDs is marginally smaller than those in the SC images, 
specifically for spots with umbral diameters larger than 30\arcsec. We obtained a 
strong positive correlation of 0.83 and 0.81 between the two quantities for the SC 
and NC images, respectively. For spots with umbral diameters less 
than 15\arcsec, the NUDs is less than 50, with a scatter amounting to about 20.

\begin{figure}[!h]
\centering
\includegraphics[width=0.45\textwidth]{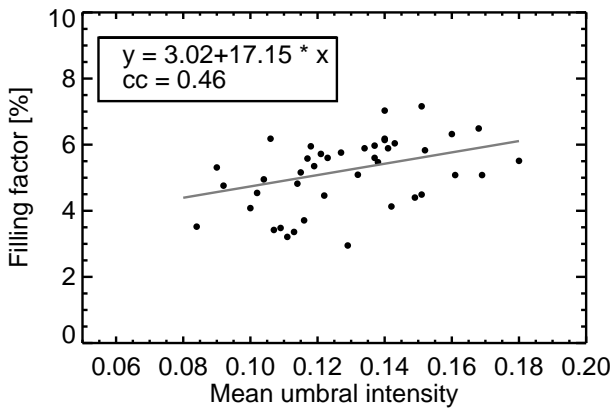}
\caption{Scatter plot between filling factor and the mean umbral intensity. The 
solid gray line is the best linear fit to the observed values. The linear correlation
coefficient between the parameters is indicated by `cc' in the inset.}
\label{fill-umbint}
\end{figure}

We also determined the fill fraction of UDs as the ratio of the total area occupied by 
UDs to the umbral area. The fill fraction ranges from 3$\%$--7.2$\%$ (bottom panels of 
Fig.~\ref{size-para}) for all sunspot umbrae.  A weak negative correlation 
coefficient of $-0.3$ and $-0.24$ was observed in the SC and NC images, respectively. 
Smaller umbrae exhibit a higher fill fraction compared to larger umbrae. In order to see
the overall trend, the average fill fraction was determined for umbral diameters within 
8\arcsec~bins and the solid line in the figure, represents the best linear fit 
to the average values. The slope of the fit indicates that the fill fraction is nearly 
independent of the umbral diameter.

A scatter plot in Figure \ref{fill-umbint} illustrates the relation between the mean 
umbral intensity and the filling factor. We observe that the filling factor shows a 
weak positive correlation with the mean umbral intensity. The filling factor increases by a factor of about 1.4 over the range of the mean umbral 
intensities. We also note that if the number of levels in the MLT routine are increased
to 45, the newly detected UDs only reflect an increase in the fill fraction of 1\% as 
compared to the case when 25 levels are used in the detection scheme.
 
\begin{figure}[]
\centering
\includegraphics[width=0.45\textwidth]{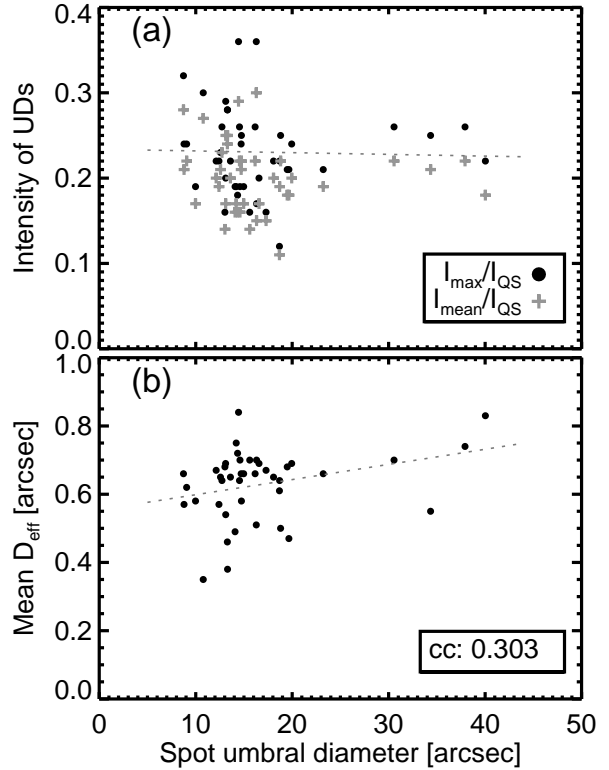}
\caption{Top: Scatter plot between UD intensity 
and umbra diameter. Filled circles and plus symbols represent the maximum and mean UD 
intensity, respectively. The dashed line represents the best linear fit between the 
maximum UD intensity and umbra diameter. Bottom: Scatter plot between effective 
diameter (D$_\mathrm{eff}$) of UDs and umbral diameter. The dashed lines represents 
the best linear fit between the umbral diameter and D$_\mathrm{eff}$.}
\label{ud_int_deffa}
\end{figure}

\begin{figure}[]
\centering
\includegraphics[width=0.45\textwidth]{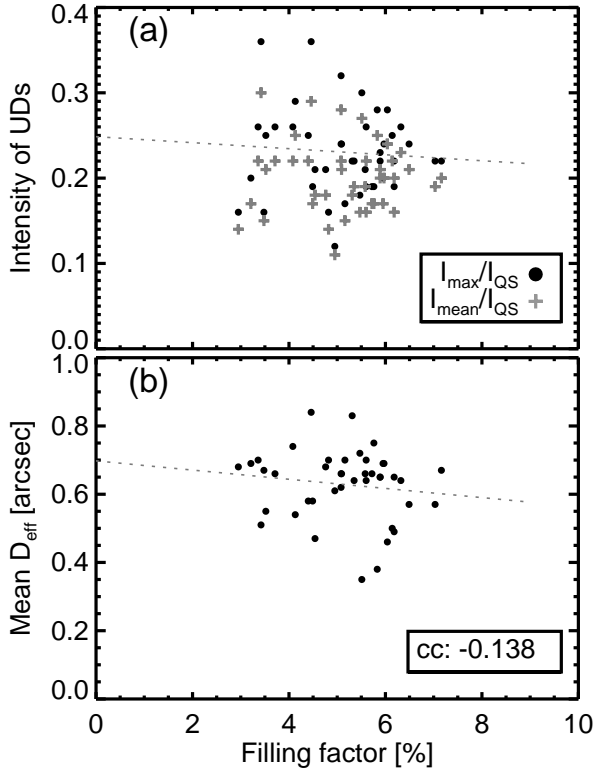}
\caption{Same as Fig.~\ref{ud_int_deffa} but with the filling factor along 
the abscissa.}
\label{ud_int_deffb}
\end{figure}

\subsection{Intensity and size of UDs}
%%%{\color{magenta}The maximum intensity and mean intensity refer to the average of the 
%%%maximum and mean intensity of all UDs in a particular umbra, respectively.} 
The top panel of Fig.~\ref{ud_int_deffa} shows that the maximum intensity of UDs in the 
dataset ranges from 0.12 to 0.36I$_\mathrm{QS}$, while the mean intensity 
of UDs varies from 0.11 to 0.3I$_\mathrm{QS}$.The mean and maximum UD intensity 
is mainly confined between 0.15--0.3I$_\mathrm{QS}$. The brightest UD in the smallest umbra 
exceeds that in the biggest umbra by a factor of 1.5. Unlike the power law relation 
between the size and intensity of the sunspot umbra (Fig.~\ref{size-para}a), neither the 
maximum nor mean intensity of UDs exhibit a correlation with the umbral size 
as seen in the top panel of Fig.~\ref{ud_int_deffa}. Figure~\ref{ud_int_deffa}(b) 
shows that the effective diameter of UDs, primarily lies 
between 0\farcs4 to 0\farcs7, with minimum and maximum values of 0\farcs35 to 
0\farcs84, respectively. A linear fit to the scatter suggests that UDs 
dwelling in smaller umbrae are smaller than those found in larger umbrae, 
with a modest correlation coefficient of about 0.3. 

We also do not observe any dependence of the intensity and effective 
diameter of UDs with the fill fraction (Fig. \ref{ud_int_deffb}). The bottom panel 
of Fig. 8 indicates that the effective diameter of UDs, is negatively 
correlated with the fill fraction, due to the large scatter in the data. 
The linear fit (Table 3) indicates that for a fill fraction of 10\% the 
effective diameter of UDs would be around 0\farcs57.

\begin{figure}[]
\centering
%%%---old---%%%\includegraphics[width=0.23\textwidth]{udimax-ibg.eps}
\includegraphics[width=0.45\textwidth]{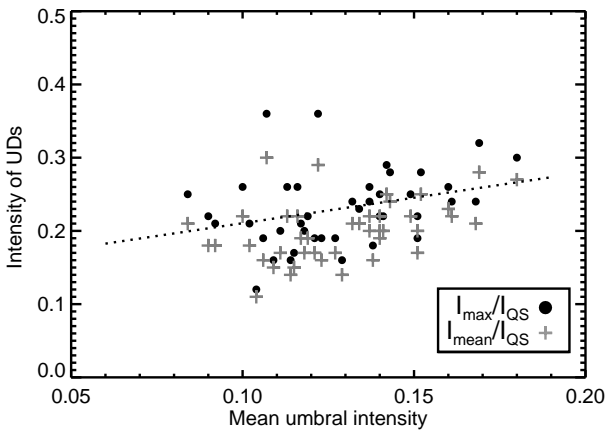}
\caption{Scatter plot showing the variation of the maximum and mean intensity 
of UDs with the mean umbral intensity observed in 42 different umbrae. 
%%%`cc' value refers to the linear correlation coefficient between parameters and 
The dashed line indicates the best linear fit between the maximum
UD intensity and mean umbral intensity.}
\label{ud-intensity}
\end{figure}

The mean umbral intensity exhibits a modest correlation with intensity
of UDs (Figure \ref{ud-intensity}). We see that the maximum intensity 
of UDs exceeds the mean umbral intensity by about 10\% with a slope of around 0.7.  

\subsection{UD parameters versus the area change and epoch of sunspots}
\begin{figure*}[!ht]
\centerline{
\includegraphics[width=0.45\textwidth]{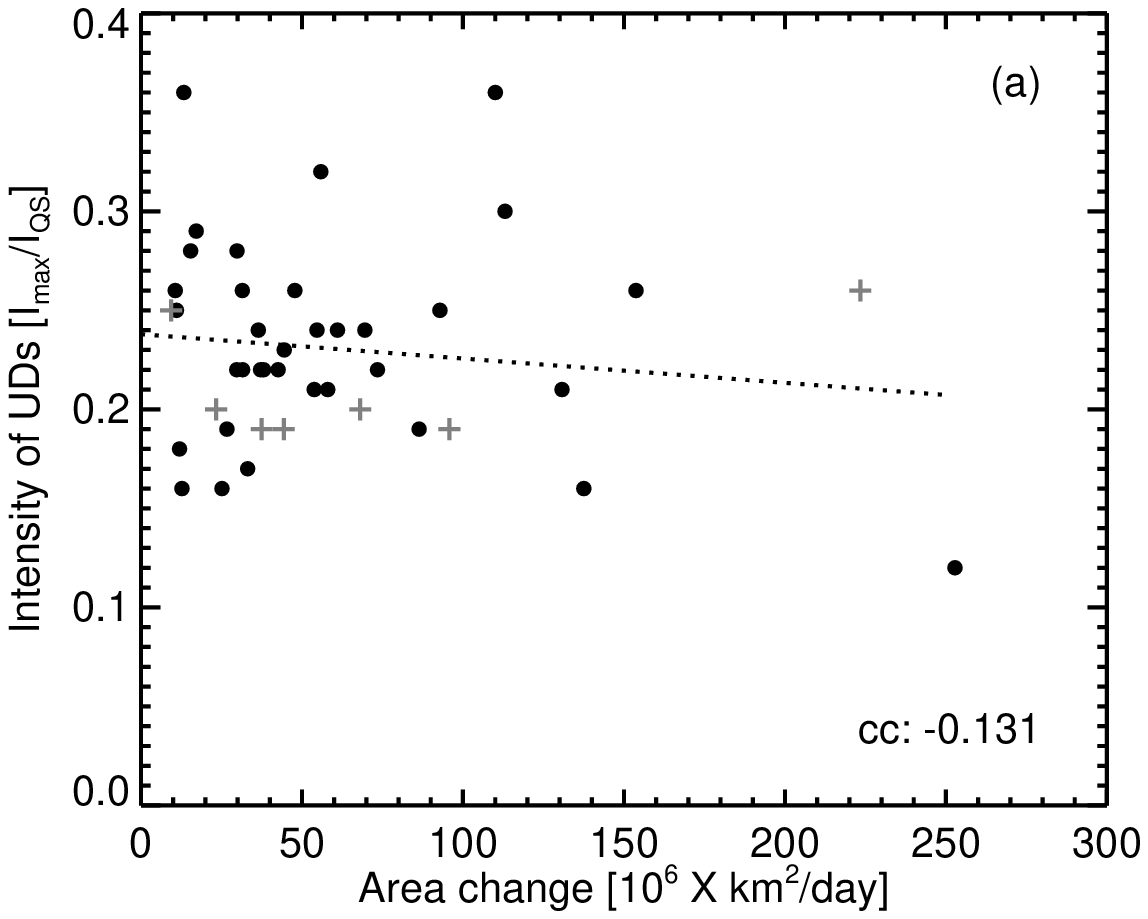}
\includegraphics[width=0.45\textwidth]{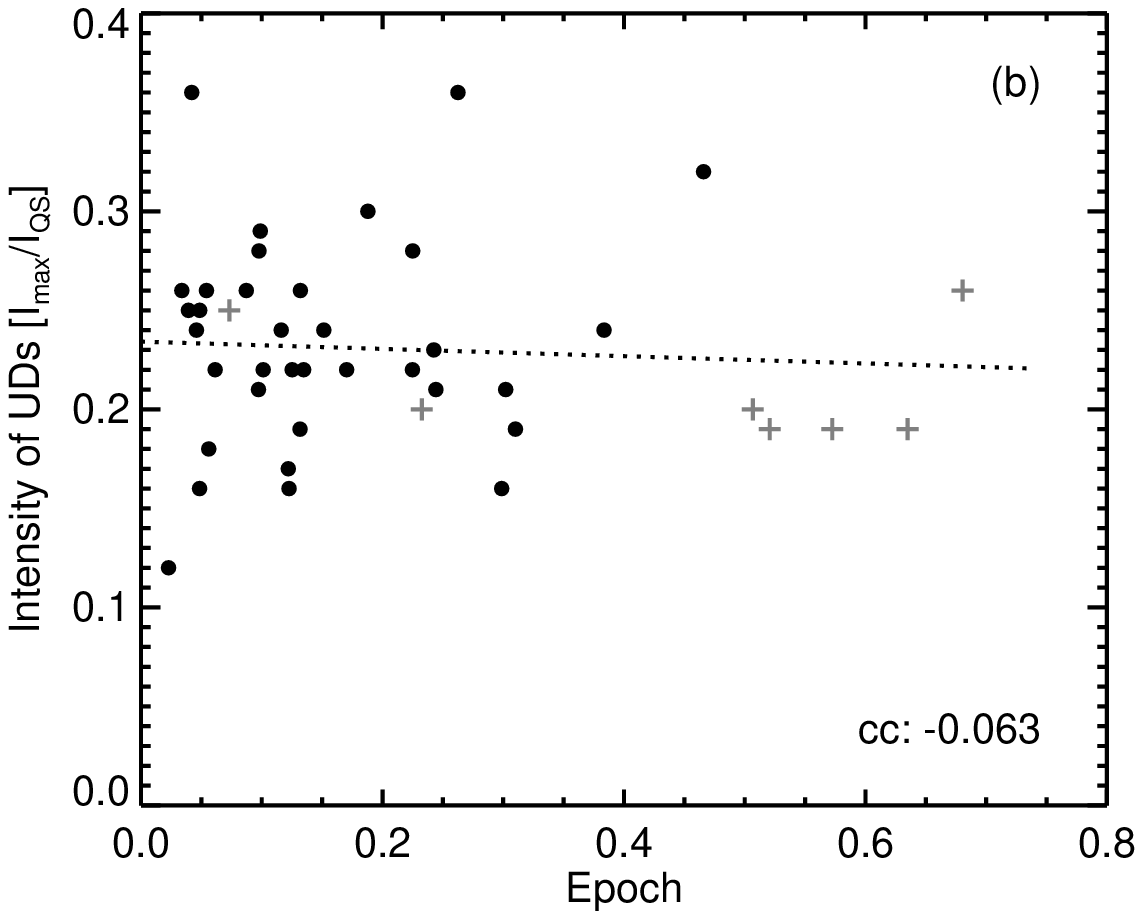}
}
\centerline{
\includegraphics[width=0.45\textwidth]{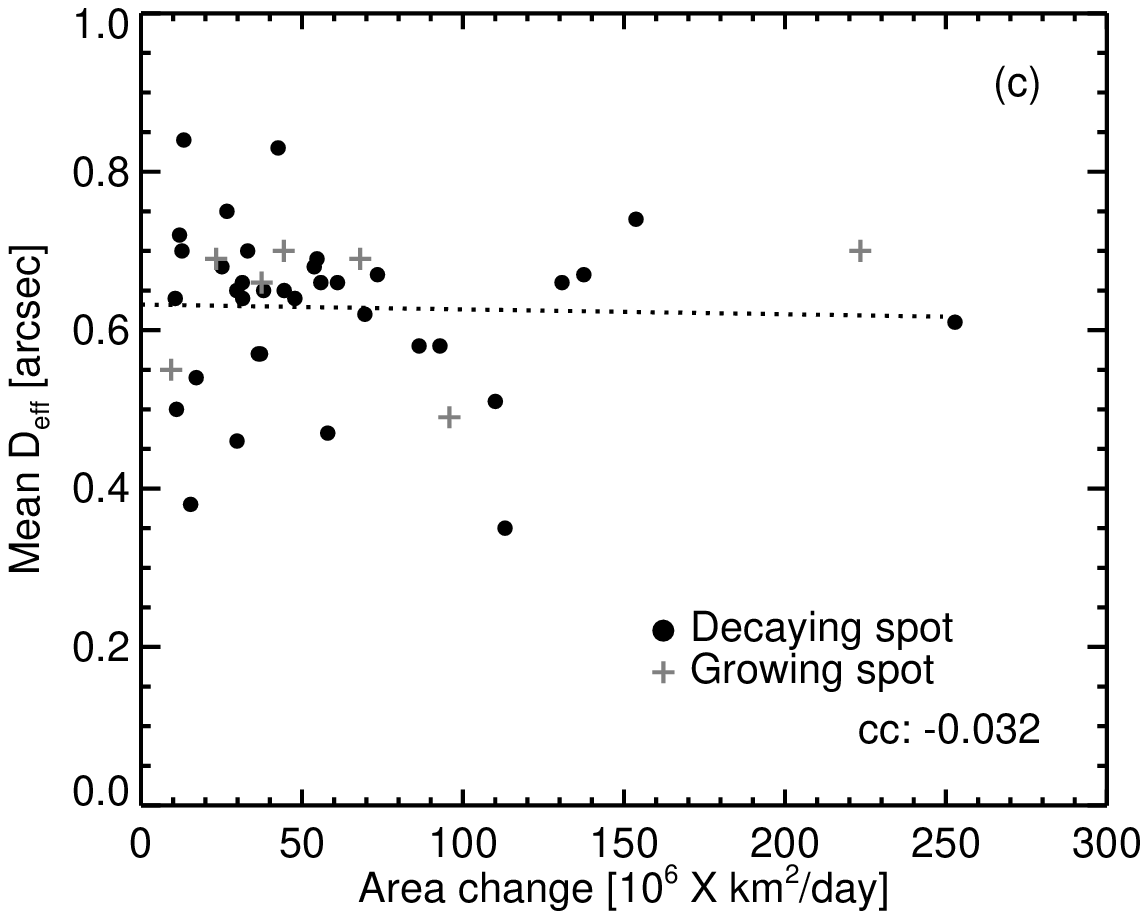}
\includegraphics[width=0.45\textwidth]{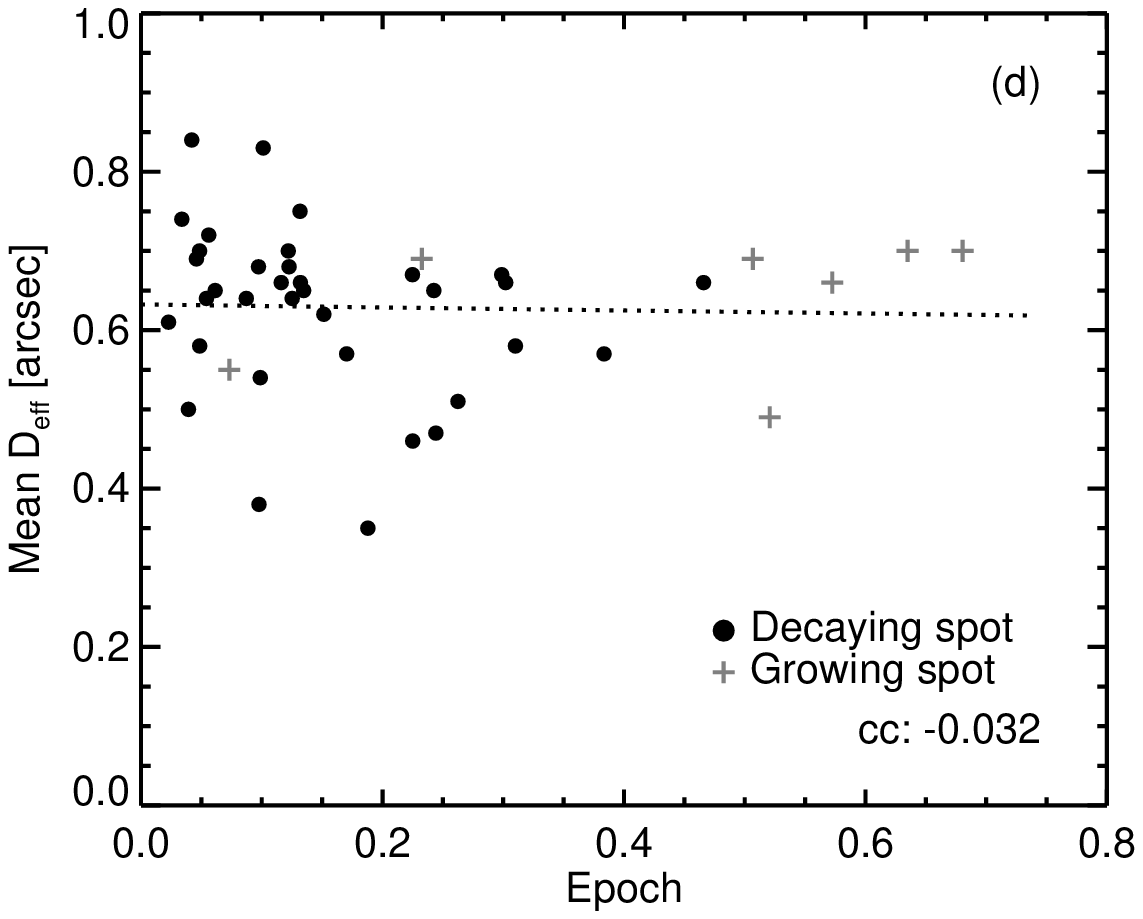}
}
\caption{Top panels: scatter plots depict the relation of maximum intensity of UDs 
with respect to the sunspot area change, i.e. decay/growth rate (left) and sunspot 
epoch (right). Bottom panels: same as the top panels but with the effective 
diameter of UDs along the ordinate. The linear correlation coefficient between 
parameters is indicated by `cc' and the dashed line denotes the best linear fit.}
\label{ud-epoch-decay}
\end{figure*}

Figure~\ref{ud-epoch-decay} (bottom panels) shows the relation of the effective diameter of UDs (D$_\mathrm{eff}$) with the decay/growth rate and epoch of 
sunspots. It is observed that the effective diameter of UDs (D$_\mathrm{eff}$) is 
insignificantly related to both the decay/groth rate and epoch for the 42 sunspots
studied here. The bottom left panel of Fig.~\ref{ud-epoch-decay}, 
indicates that a majority of the UDs are 
associated with sunspots having a slow rate of area change ($<$ 50~Mkm$^2$). Similarly, the bottom right panel of Fig.~\ref{ud-epoch-decay} illustrates that the effective diameter of UDs is also independent of the epoch of sunspots.

Similar to the effective diameter of UDs, the top panels of Fig.~\ref{ud-epoch-decay} 
demonstrate that the maximum intensity of UDs does not exhibit any trend with the
rate of area change and the epoch of sunspots. Even though UDs tend to be brighter when 
sunspots decay slowly, this variation is less than 5\%. The same is valid for the sunspot 
epoch where the UDs are only a fraction brighter during the early phase 
of sunspots.

\begin{table}[!h]
\caption{Fit parameters for the power law (y = A $\times$ x$^{B}$; 
Figures~\ref{size-para}(a -- b)) and linear (y = A + Bx; 
Figures~\ref{size-para}(c, d, e, f), \ref{fill-umbint}, \ref{ud_int_deffa}, 
\ref{ud_int_deffb}, \ref{ud-intensity}) functions. Abbreviation `ff', 
`I$_\mathrm{umb}$', and `I$_\mathrm{UD}$' correspond to the filling factor, quiet-sun normalized 
mean umbral intensity and maximum UD intensity, respectively. In the sixth column 
`cc' refers to the linear Pearson correlation coefficient between parameters. The 
numbers in the parenthesis denote one-sigma errors.}
\label{table_fitparam}
{\renewcommand{\arraystretch}{1.3}
\normalsize
\scalebox{0.75}{
\begin{tabular}{cccrcr}
\hline
Figure & x  & y & A & B & CC\\
\hline
\ref{size-para}(a)$^{\dagger} $     &D$_\mathrm{umb}$ & I$_\mathrm{umb}$    & 0.38 (0.48)       & $-0.40$ (0.45)    & $-0.69$  \\
\ref{size-para}(b)$^{\ddagger}$     &D$_\mathrm{umb}$ & I$_\mathrm{umb}$    & 0.46 (0.58)       & $-0.36$ (0.45)    & $-0.78$  \\
\ref{size-para}(c)$^{\dagger}$      &D$_\mathrm{umb}$ & NUD          & $-9.2$ (15.70)    & 3.44 (0.65)       & 0.83   \\
\ref{size-para}(d)$^{\ddagger}$     &D$_\mathrm{umb}$ & NUD          & $-11.52$ (14.41)  & 3.02 (0.60)       & 0.81   \\
\ref{size-para}(e)$^{\dagger}$      &D$_\mathrm{umb}$ & ff           & 5.77 (1.03)       & $-0.06$ (0.046)   & $-0.30$  \\
\ref{size-para}(f)$^{\ddagger}$     &D$_\mathrm{umb}$ & ff           & 5.30 (0.80)       & $-0.05$ (0.04)    & $-0.24$  \\
\ref{fill-umbint}$^{\dagger}$       &I$_\mathrm{umb}$ & ff           & 3.02 (0.71)       & 17.15 (5.06 )     & 0.46       \\
\ref{ud_int_deffa}(a)$^{\dagger}$   &D$_\mathrm{umb}$ & I$_\mathrm{UD}$     & 0.23 (0.02)       & $-2.01 (1.14)\times 10^{-4}$ & $-0.028$ \\
\ref{ud_int_deffa}(b)$^{\dagger}$   &D$_\mathrm{umb}$ & D$_\mathrm{eff}$    & 0.55 (0.04)       & $4.44 (2.21)\times 10^{-3}$  & $0.30$  \\
\ref{ud_int_deffb}(a)$^{\dagger}$   & ff       & I$_\mathrm{UD}$     & 0.25 (0.04)       & $-3.5 (7.6)\times 10^{-3}$   & $-0.073$ \\
\ref{ud_int_deffb}(b)$^{\dagger}$   & ff       & D$_\mathrm{eff}$    & 0.70 (0.08)       & $-0.013$ (0.015)  & $-0.138$ \\
\ref{ud-intensity}$^{\dagger}$ 	    &I$_\mathrm{umb}$ & I$_\mathrm{UD}$     & 0.14 (0.04)       & 0.70 (0.33)       & 0.31       \\
\ref{ud-epoch-decay}(a)$^{\dagger}$ & $\mathrm{dA/dt}$  & I$_\mathrm{UD}^{\mathrm{max}}$ &  0.24 (0.02)    & $-$1.2 (1.4) $\times 10^{-4}$  & $-0.131$ \\
\ref{ud-epoch-decay}(b)$^{\dagger}$ & Epoch    & I$_\mathrm{UD}^{\mathrm{max}}$ &  0.23 (0.01)    & $-$1.8 (4.6) $\times 10^{-4}$  & $-0.063$ \\
\ref{ud-epoch-decay}(c)$^{\dagger}$ & $\mathrm{dA/dt}$  & D$_\mathrm{eff}$    &  0.63 (0.02)      & $-$6.13 (3.0) $\times 10^{-5}$ & $-0.032$ \\
\ref{ud-epoch-decay}(d)$^{\dagger}$ & Epoch    & D$_\mathrm{eff}$    &  0.63 (0.02)      & $-$1.9 (9.3) $\times 10^{-4}$  & $-0.032$ \\
\hline
\end{tabular}}
}
\begin{flushleft}
The parameters retrieved from SC and NC images are indicated 
by `${\dagger}$' and `${\ddagger}$' superscript, respectively. The rate of area change
is denoted by $\mathrm{dA/dt}$.
\end{flushleft}
\end{table}

\section{Discussion}
\label{discuss}
Umbral dots are manifestations of small-scale magnetoconvection in sunspots. Our study 
shows that there is a dependence of the mean umbral brightness with the spot size which 
is in agreement with \cite{2007A&A...465..291M}, where smaller spots are brighter than 
larger ones. This would imply that darker spots comprise stronger magnetic fields which 
suppress magnetoconvection as suggested by
\cite{1977SoPh...55..335M,1992SoPh..141..253K,2002SoPh..207...41L}. Thus, if umbral dots 
are manifestations of small-scale, magnetoconvection, then one would expect that their 
physical properties have a bearing on the macro-properties of a sunspot. In the context 
of Parker's ``jelly-fish'' model \citep{1979ApJ...234..333P}, the energy transport in 
umbrae ought to be more vigorous during the late phase of sunspots as they approach
fragmentation, and the reduced magnetic pressure would be insufficient to overcome 
the resulting gas pressure. This should be seen as brighter and/or larger umbral dots, 
that could then be used as a proxy for several large-scale properties of sunspots, 
namely, their area, rate of decay, and evolutionary phase. With this motivation in mind, 
we have selected a large set of sunspots and tracked them during their transition across 
the solar disc. By employing high resolution G-band filtergrams from {\em Hinode}, close 
to disc center, the physical properties of umbral dots were estimated and related to 
the above mentioned sunspot properties. We have taken care of instrumental stray-light 
in the {\em Hinode} filtergrams, which is known to affect the geometrical and photometric
properties of umbral dots \citep{2012ApJ...752..109L} and our analysis primarily focuses on the intensity and effective diameter. 

We find that a strong linear relationship exists between the number of umbral dots and 
umbral diameter, which would suggest that in bigger spots there 
are larger spaces for convection 
to occur within the umbra. However, the fill fraction of umbral dots is nearly independent 
of the umbral diameter and accounts for less than 10\% of the umbral area, which is in 
agreement with \cite{1993ApJ...415..832S,2005A&A...442..323S}. This stems from the fact 
that the diameter of umbral dots does not show any variation with the spot size. In 
addition, both the mean and maximum intensity as well as diameter 
of umbral dots do not show any visible trend 
with the area decay rate and the spot epoch either, exhibiting very weak, negative 
correlations. We find, that although UDs tend to be brighter during 
the late phase of sunspots, this variation is less than 5\%. We also 
observed a similar behavior with the rate of area change. We also see that UDs tend to 
be smaller for spots that are either in an advanced stage of evolution or decay faster. 
This is reflected in a negative, although very weak, correlation coefficient. Our 
results show that the maximum intensity of UDs is about 10\% brighter than the mean 
umbral intensity. We interpret this as a combined effect of the small fill fraction 
of less than 10\% and the weak dependence between the diameter
of UDs and the host umbra. This would indicate that the dependence of umbral 
intensity with the spot size originates primarily from the background regions 
of the umbra and to a very small fraction from UDs. The lack of a relationship 
between the properties of umbral dots and the macro-properties of the parent 
spots is discussed below.

In addition to umbral dots, light bridges represent large-scale, convective intrusions 
in the umbrae of sunspots and pores 
\citep{1979SoPh...61..297M,1994ApJ...426..404S,2004SoPh..221...65L,2008SoPh..252...43L,
2009ApJ...704L..29L}. The association between these two phenomena has been established 
in several works 
\citep{1987SoPh..112...49G,2002A&A...383..275H,2008ApJ...672..684R}. 
\cite{2007PASJ...59S.577K} studied the formation of sunspot light bridge using 
{\em Hinode} observations. They 
found that the formation was preceded by an inward motion of umbral dots that appeared 
well within the umbra and not the penumbra. They interpreted this observation as a sign 
of the weakening of the magnetic field by the hot rising plasma that then allowed several 
umbral dots from the leading edges of penumbral filaments to intrude further into the 
umbra, forming a light bridge out of a collection of umbral dots. The formation of a light 
bridge is associated with a rapid increase of intensity, from umbral to penumbral values 
in about 4~hr, which is accompanied by a large reduction in the field strength
\citep{2012ApJ...755...16L}. In light of the above, the lack of trend between 
the umbral dot size and the spot area suggests that the interaction of the magnetic 
field and convection, within the umbra, occurs over a set of distinct, interchangeable 
spatial scales. Thus, depending on the underlying conditions of the magnetic and gas 
pressure, umbral dots would coalesce to form light bridges during late stages of spot 
decay and light bridges would disintegrate into umbral dots during the spot's 
maturity \cite[see for example Fig.2 of][]{2010A&A...512L..1S}. Since the contribution 
of light bridges to the intensity of umbral dots was excluded in our analysis, it 
would be necessary to determine how these structures influence the size and intensity 
of the latter.

Another possibility of the lack of any discernible relation between the intensity and 
diameter of umbral dots with the spot size, could be attributed to the shallow depth 
to which these structures extend to. MHD simulations of magnetocovnection in the umbra 
by \cite{2006ApJ...641L..73S}, show that umbral dots correspond to vertically rising 
convective plumes from a depth of around 2~Mm, while light bridges extend a little 
deeper to around 6--7~Mm \citep{2010ApJ...720..233C}. Even with this simulation 
depth, the photospheric properties of light bridges are in good agreement with 
those typically seen in observations. These depths only represent a small fraction 
of the solar convection zone which is strongly stratified by 6 orders of magnitude 
over a depth of 200~Mm \citep{1991ApJ...378..413C}. This poses an enormous computational
challenge, and as such numerical models focus either on the deep convection zone leaving 
out the uppermost 10--20~Mm or on the uppermost 10~Mm including the solar photosphere. 

According to \cite{2005A&A...441..337S}, during the final phase of the ascent of a 
rising flux loop towards the surface, the upper part of the loop develops a buoyant 
upflow of plasma. The combination of the pressure build-up by the upflow and the 
cooling of the upper layers of an emerged flux tube by radiative losses at the 
surface leads to a progressive weakening of the magnetic field at depths of several 
Mm, which can lead to a dynamic disconnection of the bipolar structure from its 
magnetic roots. The disconnection depth extends to a few tens of Mm as shown by 
\cite{2009A&A...506..875S}, but is associated with only one-third of the sample 
analyzed. Similar values of the disconnection depth have been reported by 
\cite{2010ApJ...714L.196M} who find a linear relation of the above with the 
remaining lifetime of active regions. This would suggest that umbral dots and 
light bridges are strongly influenced by near-surface convective flows, rather than 
those which are associated with the severing of sunspots from their roots, which 
occur much deeper.

\section{Conclusions}
\label{conclu}
The study of umbral dots with high-resolution data is crucial for understanding 
small-scale magnetoconvection in sunspot umbrae, that can constrain existing 
sunspot models in a more robust manner. In this article, we have attempted to 
relate the physical properties of umbral dots with the large-scale properties 
of sunspots, in order to determine if the underlying physical processes, that 
influence the evolution and stability of the latter, are indeed scale-invariant. 
We do not find any significant relationship between the effective diameter
of umbral dots with the sunspot area, epoch, and decay rate. The same is observed
with the mean and maximum intensity of the umbral dots.
We conclude that the above could either be due to the distinct transition of spatial 
scales associated with overturning convection in the umbra, where umbral dots can 
coalesce to form light bridges, or the shallow depth associated with umbral dots 
which make them impervious to the deeper, large-scale, convective flows that affect 
the anchoring of sunspots, or both of the above. We intend to 
investigate if this lack of trend is 
extended to sunspots over an entire solar cycle and with spots that 
have an umbral radius greater than 25\arcsec. Facilities such as the 
1.5-m GREGOR solar telescope \citep{2012AN....333..796S} and the 4-m Daniel K. Inoyue 
Solar Telescope \cite[DKIST, formerly ATST;][]{2003SPIE.4853..240K} will 
be extremely important in these investigations as spatially resolved UDs
would allow improved statistics by providing evidence of magnetoconvection
at the smallest spatial scales.
\\
\noindent {\bf Acknowledgements:} 
{\em Hinode} is a Japanese mission developed and launched by ISAS/JAXA, collaborating with 
NAOJ as a domestic partner, NASA and STFC (UK) as international partners. Scientific 
operation of the {\em Hinode} mission is conducted by the {\em Hinode} science team 
organized at ISAS/JAXA. This team mainly consists of scientists from institutes in 
the partner countries. Support for the post-launch operation is provided by JAXA and 
NAOJ (Japan), STFC (U.K.), NASA, ESA, and NSC (Norway). The HMI data used here are 
courtesy of NASA/SDO and the HMI science team. The Center of Excellence in Space 
Sciences India is funded by the Ministry of Human Resource Development, Government 
of India. 
REL is grateful for the financial assistance from SOLARNET--the European Commission's 
FP7 Capacities Programme under Grant Agreement number 312495. This work was also 
supported by grants AYA2014-60476-P and SP2014-56169-C6-2-R at the Instituto de 
Astrof\'isica de Canarias, Tenerife, Spain. We thank the referee for reviewing 
our article and for providing insightful comments.

\bibliographystyle{aasjournal}
\bibliography{new-ref}  

\end{document}